\documentclass[11pt,a4paper]{article}
\usepackage{amsmath,amssymb}
\usepackage{epsfig,graphicx}

\begin{document}

\title{\bf Nature of Light Scalar Mesons in
Bright Light of Photon-Photon Collisions}
\author{N.~N.~Achasov\footnote{{\bf e-mail}: achasov@math.nsc.ru},
G.~N.~Shestakov\footnote{{\bf e-mail}: shestako@math.nsc.ru}
\\
\small{\em Sobolev Institute for Mathematics, Academician Kopiug
Prospect, 4, 630090, Novosibirsk, Russia}}
\date{}
\maketitle

\begin{abstract}
The surprising thing is that the light scalar meson problem, arising
50 years ago from the linear sigma model (LSM) with spontaneously
broken chiral symmetry, has become central in the nonperturbative
quantum chromodynamics (QCD), because it has been made clear that
LSM could be the low energy realization of QCD. First, we review
briefly signs of  four-quark nature of  light scalars. Then we show
that  the light scalars are produced in the two-photon collisions
via four-quark transitions in contrast to the classic $P$ wave
tensor $q\bar q$ mesons that are produced via two-quark transitions
$\gamma\gamma\to q\bar q$. Thus we get new evidence of the
four-quark nature of the lower scalar states.
\end{abstract}

\hspace*{0.4cm}{\bf Outline}
\begin{enumerate}
\item{} Introduction
\item{} Evidence for the four-quark nature of light scalar
mesons \\
i)\ \  Normal ($q\bar q$) and inverted ($q^2\bar q^2$) mass spectra
\\
ii)\  The $\phi(1020)$ meson radiative decays about light scalars\\
iii) Chiral shielding of the $\sigma(600)$ meson in
$\pi\pi\to\pi\pi$
\item{} Light scalar manifestations in
$\gamma\gamma$ collisions \\ i)\ \  Prediction of the four-quark
model. New stage of high statistics measurements,\\ \hspace*{0.5cm}the Belle data\\
ii)\ \,Dynamics of the $\sigma(600)$ and $f_0(980)$ production in
$\gamma\gamma\to\pi\pi$ \\
iii) Dynamics of the  $a_0(980)$ production in
$\gamma\gamma\to\pi^0\eta$
\item{} Future trends: the $\sigma(600)$, $f_0(980)$ and
$a_0(980)$ investigations in $\gamma\gamma\to K\bar K$ and\\ in
$\gamma\gamma^*$ collisions
\item{} Summary
\end{enumerate}

{\bf 1. Introduction.}

The scalar channels in the region up to 1 GeV became a stumbling
block of QCD. The point is that both perturbation theory and sum
rules do not work in these channels because there are not solitary
resonances in this region.

At the same time the question on the nature of the light scalar
mesons is major for  understanding the mechanism of the chiral
symmetry realization, arising from the confinement, and hence for
understanding the confinement itself.\\

{\boldmath \bf 2. Evidence for the four-quark nature of light scalar
mesons.

i) \,Normal ($q\bar q$) and inverted ($q^2\bar q^2$) mass spectra.}

\begin{figure}\begin{center}
\includegraphics[width=100mm]{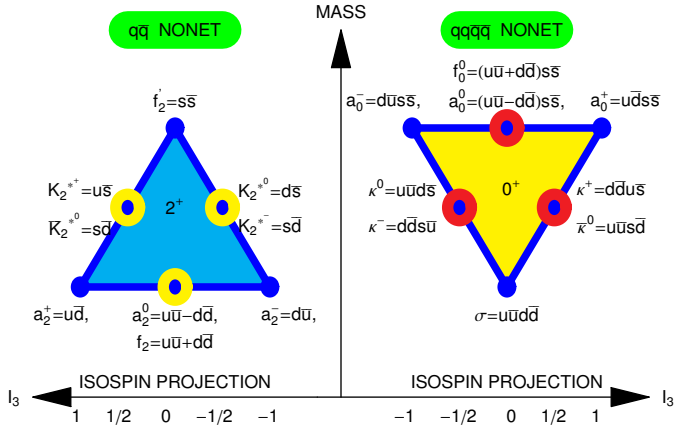} \vspace{-0.2cm}
\caption{{\footnotesize Normal $2^{++}$ ($q\bar q$) and inverted
$0^{++}$ ($q^2\bar q^2$) mass spectra.}}\label{MassSpectra}
\end{center}\end{figure}

The mass spectrum of the light scalars $\sigma(600)$, $\kappa(800)$,
$a_0(980)$, and $f_0\,(980)$ gives an idea of their $q^2\bar q^2$
structure. Really, this scalar nonet turns out to be inverted in
comparison with the classical $P$ wave $q\bar q$ tensor meson nonet
(Fig. \ref{MassSpectra}). In the naive quark model such a nonet
cannot be understood as the $P$ wave $q\bar q$ nonet, but it can be
easy understood as the $S$ wave $q^2\bar q^2$ nonet, where
$\sigma(600)$ has no strange quarks, $\kappa(800)$ has the $s$
quark, and $a_0(980)$ and $f_0(980)$ have the $s\bar s$ pair.

\begin{figure}\begin{center}\vspace{-1.1cm}
\includegraphics[height=60mm]{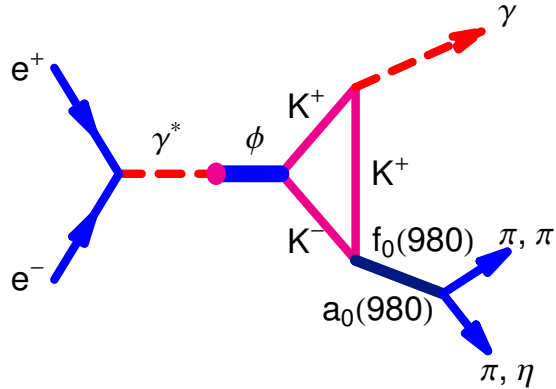}\vspace{-0.4cm}
\caption{{\footnotesize $K^+K^-$ loop mechanism of
$f_0(980)/a_0(980)$ production in $\phi(1020)\to\gamma[f_0(980)
/a_0(980)]$ decays.}}\label{KK-loop}
\end{center}\end{figure}

{\boldmath \bf ii) The $\phi(1020)$ meson radiative decays about
light scalars \cite{AI89,AK03,AK06}.}

\begin{figure}\begin{center}\begin{tabular}{cc}
\includegraphics[height=1.35in]{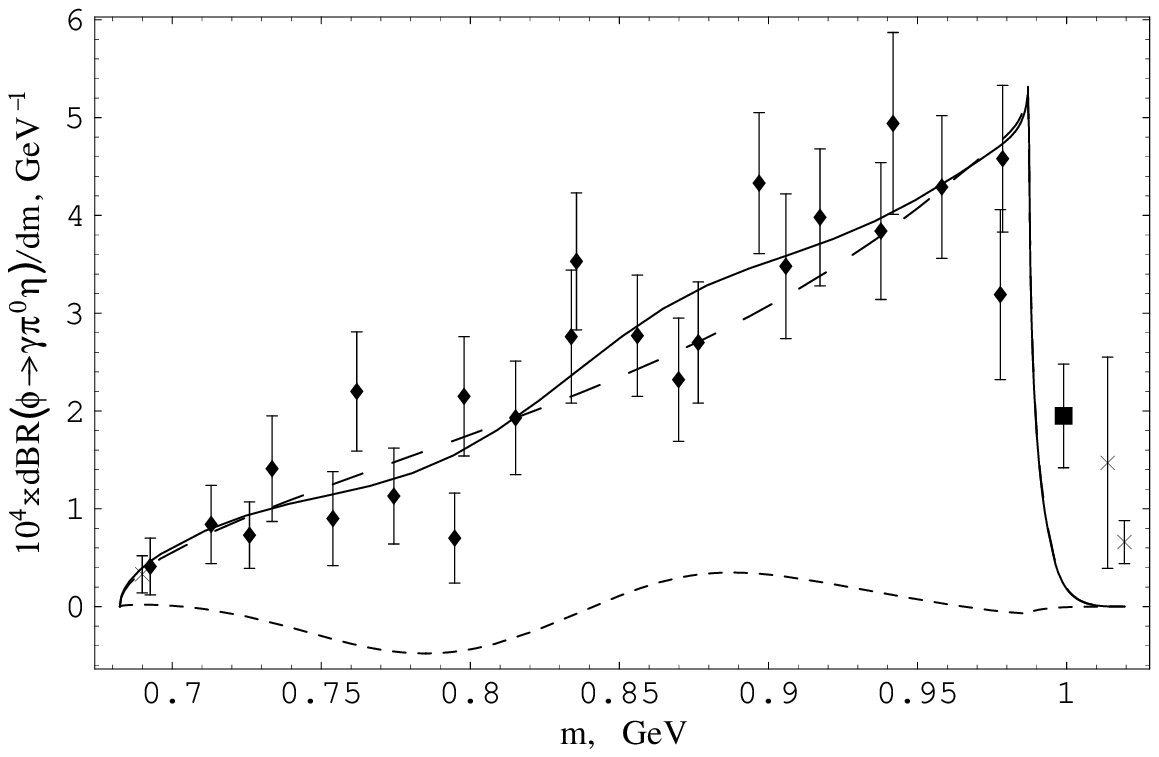}\
\includegraphics[height=1.35in]{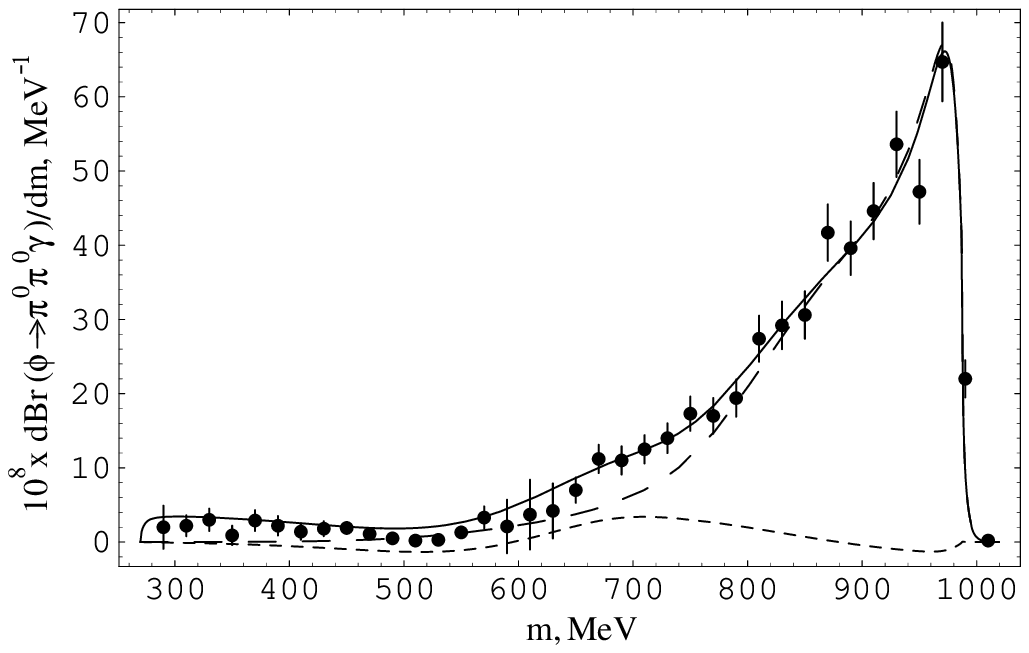}
\end{tabular}\\
\vspace{-0.3cm}\caption{{\footnotesize The left and right plots
illustrate the fits \cite{AK03,AK06} to the KLOE data for the
$\pi^0\eta$ and $\pi^0\pi^0$ mass spectra in the
$\phi$\,$\to$\,$\gamma\pi^0\eta$ \cite{Al02a} and
$\phi$\,$\to$\,$\gamma \pi^0\pi^0$ \cite{Al02b} decays,
respectively.}}\label{KLOE-spectra}\end{center}\end{figure}

At the end of eighties it was shown that the study of the radiative
decays $\phi\to\gamma f_0\to \gamma\pi\pi$ and $\phi\to\gamma
a_0\to\gamma\pi\eta$ can shed light on the problem of $f_0(980)$ and
$a_0(980)$ mesons \cite{AI89}. Now these decays have been studied
not only theoretically but also experimentally with the help of the
SND and CMD-2 detectors at Budker Institute of Nuclear Physics in
Novosibirsk and the KLOE detector at the DA$\Phi$NE $\phi$-factory
in Frascati. When basing the experimental investigations
\cite{AI89}, it was  suggested  the kaon loop model
$\phi$\,$\to$\,$K^+K^-$\,$\to$\,$\gamma f_0(980)$\,$\to$\,$
\gamma\pi\pi$  and $\phi$\,$\to$\,$K^+K^-$\,$\to$\,$\gamma
a_0(980)$\,$\to$\,$ \gamma\pi^0\eta$ (Fig. \ref{KK-loop}). This
model is used in the data treatment and is ratified by experiment,
see (Fig. \ref{KLOE-spectra}). Both intensity and mechanism of the
$f_0(980)$ and $a_0(980)$ production in the radiative decays of the
$\phi(1020)$, via the $q^2\bar q^2$ transitions $\phi\to
K^+K^-\to\gamma[f_0(980)/a_0(980)]$, testify to their $q^2\bar q^2$
nature \cite{AI89,AK03,AK06}.

{\boldmath \bf iii) Chiral shielding of the $\sigma(600)$ meson in
$\pi\pi$\,$\to$\,$\pi\pi$ and other evidence \cite{AS94,AS07,
Ac08}.}

Hunting the light $\sigma$ and $\kappa$ mesons had begun in the
sixties already. But long-standing unsuccessful attempts to prove
their existence in a conclusive way entailed general disappointment
and an information on these states disappeared from PDG Reviews. One
of principal reasons against the $\sigma$ and $\kappa$ mesons was
the fact that both $\pi\pi$ and $\pi K$ scattering phase shifts do
not pass over $90^0$ at putative resonance masses. Situation has
changed when we showed that in the $SU(2)\times SU(2)$ linear
$\sigma$ model there is a negative background phase which hides the
$\sigma$ meson \cite{AS94,AS07,Ac08}. It has been made clear that
shielding wide lightest scalar mesons in chiral dynamics is very
natural. This idea was picked up and triggered new wave of
theoretical and experimental searches for the $\sigma$ and $\kappa$
mesons.

\begin{figure}\begin{center}\vspace{-0.2cm}
\includegraphics[width=95mm]{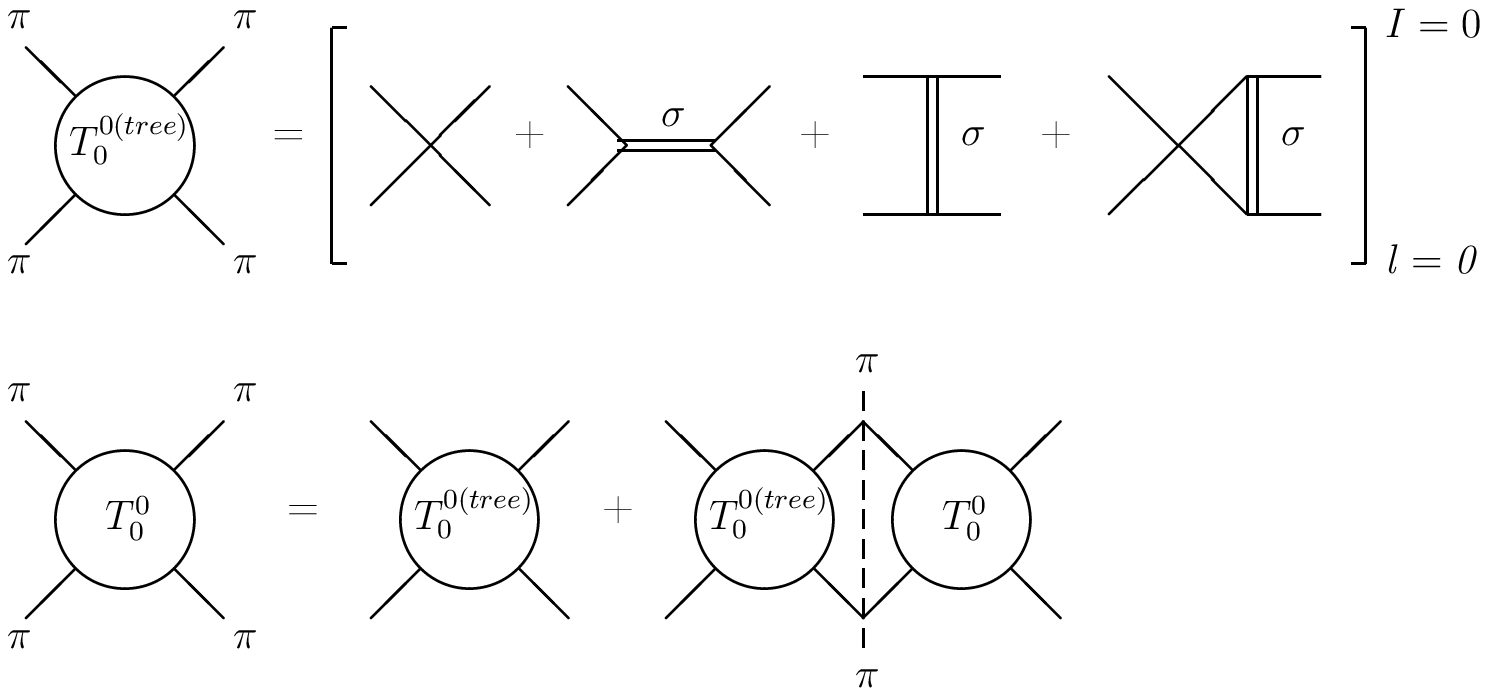}
\vspace{-0.3cm}\caption{{\footnotesize The graphical representation
of the $S$ wave $I=0$ $\pi\pi$ scattering amplitude $T^0_0$.
}}\label{ChiShi-1}\end{center}\end{figure}

\begin{figure}\begin{center}\vspace{-0.2cm}
\includegraphics[height=40mm,width=110mm]{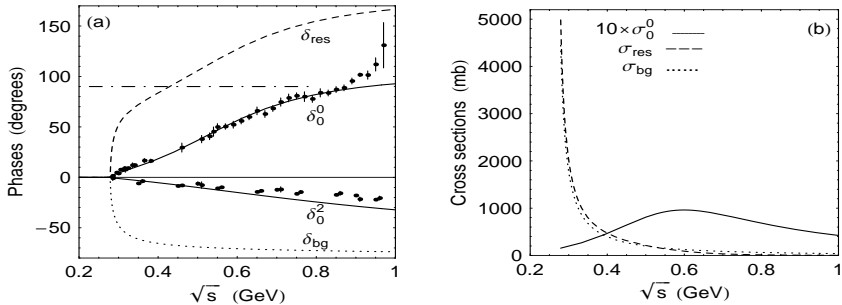}
\vspace{-0.3cm}\caption{{\footnotesize The $\sigma$ model. Our
approximation. Chiral shielding of the
$\sigma(600)$.}}\label{ChiShi-2}\end{center}\end{figure}

Chiral shielding of the $\sigma(600)$ can be easily revealed with
the use of the $S$ wave $I=0$ $\pi\pi$ scattering amplitude $T^0_0$
satisfying the simplest Dyson equation with the real $\pi$ mesons in
the intermediate state (Fig. \ref{ChiShi-1}). It is illustrated in
Fig. \ref{ChiShi-2}(a) with the help of the $\pi\pi$\,$\to$\,$\pi
\pi$ phase shifts $\delta_{res}$, $\delta_{bg}$, $\delta^0_0=
\delta_{res}+\delta_{bg}$ and in Fig. \ref{ChiShi-2}(b) with the
help of the corresponding cross sections. Note that in the $\sigma$
meson propagator $$ \frac{1}{D_\sigma (s)}=\frac{1}{M^2_{res}-s
+\mbox{Re}\Pi_{res}(M^2_{res})- \Pi_{res}(s)}\,,$$ the $\sigma$
self-energy $\Pi_{res}(s)$ is caused by the intermediate $\pi\pi$
states, that is, by the four-quark intermediate states. This
contribution shifts the Breit-Wigner (BW) mass greatly
$m_\sigma-M_{res}$\,=\,0.50\,GeV. So, half the BW mass is determined
by the four-quark contribution at least. The imaginary part
dominates the propagator modulus in the region
300\,MeV\,$<\sqrt{s}<$\,600\,MeV. So, the $\sigma$ field is
described by its four-quark component at least in this energy
(virtuality) region.\\

{\boldmath \bf 3. Light scalar manifestations in $\gamma\gamma$
collisions \cite{AS07,Ac08,ADS82,AS88,AS05,AS08,AS10}.

i) \,Prediction of the four-quark model. New stage of high
statistics measure-

$\mbox{\ \ \ \,}$ments, the Belle data \cite{Mo07,Ue08,Ue09}.}

\begin{figure}\begin{center}\vspace{-1cm}
\includegraphics[height=65mm]{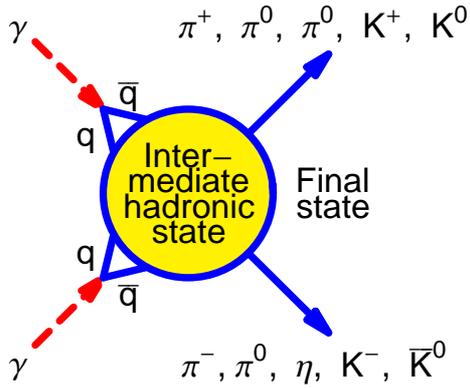}\vspace{-0.3cm}
\caption{{\footnotesize Probe for the quark structure of light
scalars.}} \label{GG-piKeta}\end{center}\end{figure}

Photons are probes of the quark structure of hadrons. Investigations
of the mechanisms of the reactions $\gamma\gamma\to\pi^+\pi^-$,
$\gamma\gamma\to\pi^0\pi^0$, $\gamma\gamma\to\pi^0\eta$,
$\gamma\gamma\to K^+K^-$, and $\gamma\gamma\to K^0\bar K^0$  (Fig.
\ref{GG-piKeta}) are an important constituent of the light scalar
meson physics. Twenty eight years ago we predicted \cite{ADS82} that
if the $a_0(980)$ and $f_0(980)$ are the $q^2\bar q^2$ MIT bag
states, then their $\gamma\gamma$ widths,
$\Gamma(a_0(980)\to\gamma\gamma)\sim\Gamma
(f_0(980)\to\gamma\gamma)\sim 0.27\,\mbox{keV}$, are an order of
magnitude smaller than those of the $q\bar q$ mesons $\eta'$,
$f_2(1270)$, and the theoretical estimates in the $q\bar q$ model.
Experiment supported this prediction: $\Gamma (a_0\to\gamma\gamma)
=(0.19\pm 0.07 ^{+0.1}_{-0.07})\, \mbox{keV, Crystal Ball,}$ $\Gamma
(a_0\to\gamma\gamma)=(0.28\pm 0.04\pm 0.1)\, \mbox{keV, JADE,}$
$\Gamma (f_0\to\gamma\gamma)=(0.31\pm 0.14\pm 0.09)\, \mbox{keV,
Crystal Ball,}$ $\Gamma (f_0\to\gamma\gamma)=(0.29\pm 0.07\pm
0.12)\, \mbox{keV, MARK II}$. When in the $q\bar q$ model it was
anticipated $\Gamma(a_0\to \gamma\gamma)\approx(1.5-5.9)\Gamma(a_2
\to\gamma\gamma)\approx (1.5-6)$\,keV, $\Gamma(f_0\to\gamma
\gamma)\approx(1.7-5.5)\Gamma (f_2\to\gamma\gamma)\approx(4.5-14)
$\,keV.

Recently, the experimental investigations have made great
qualitative advance. The Belle Collaboration  published data on
$\gamma\gamma\to\pi^+\pi^-$ \cite{Mo07}, $\gamma\gamma\to\pi^0\pi^0$
\cite{Ue08}, and $\gamma\gamma\to\pi^0\eta$ \cite{Ue09}, whose
statistics are huge. They not only proved the theoretical
expectations based on the four-quark nature of the light scalar
mesons,  but also have allowed to elucidate the principal mechanisms
of these processes. Specifically, the direct coupling constants of
the $\sigma(600)$, $f_0(980)$, and $a_0(980)$ resonances with the
$\gamma\gamma$ system are small and their decays into photons are
the four-quark transitions caused by the rescattering mechanisms $
\sigma$\,$\to $\,$\pi^+\pi^-$\,$\to$\,$\gamma\gamma$, $
f_0(980)/a_0(980)$\,$\to$\,$K^+K^-$\,$\to$\,$\gamma\gamma$, and
$a_0(980)$ $\to$\,$\pi^0\eta$\,$\to$\,$\gamma\gamma$, in contrast to
the two-photon decays of the classic $P$ wave tensor $q\bar q$
mesons $a_2(1320)$, $f_2(1270)$ and $f'_2(1525)$, which are caused
by the direct two-quark transitions $q\bar q$\,$\to
$\,$\gamma\gamma$ in the main.

As a result the practically model-independent prediction of the
$q\bar q$ model for the $2^{++}\gamma\gamma$ coupling constants $
g^2_{f_2\gamma\gamma}:g^2_{a_2\gamma\gamma}=25:9$ agrees with
experiment rather well; $\Gamma_{f_2\to\gamma\gamma}\approx2.8$ keV,
$\Gamma_{a_2\to\gamma\gamma}\approx1$ keV. The two-photon light
scalar widths averaged over resonance mass distributions are:
$\langle\Gamma_{f_0\to\gamma\gamma}\rangle_{\pi\pi}$\,$\approx
$\,0.19 keV \cite{AS05,AS08}, $\langle\Gamma_{a_0\to\gamma
\gamma}\rangle_{\pi\eta}$\,$\approx$\,0.4 keV \cite{AS10}, and
$\langle\Gamma_{\sigma \to\gamma\gamma }\rangle_{\pi\pi}$\,$\approx
$\,0.45 keV \cite{AS08}. As to the ideal $q\bar q$ model prediction
for the $0^{++}$\,$\to$\,$\gamma\gamma$ coupling constants
$g^2_{f_0\gamma\gamma}:g^2_{a_0\gamma\gamma}=25:9$, it is excluded
by experiment.

\begin{figure}\begin{center}
\includegraphics[height=60mm]{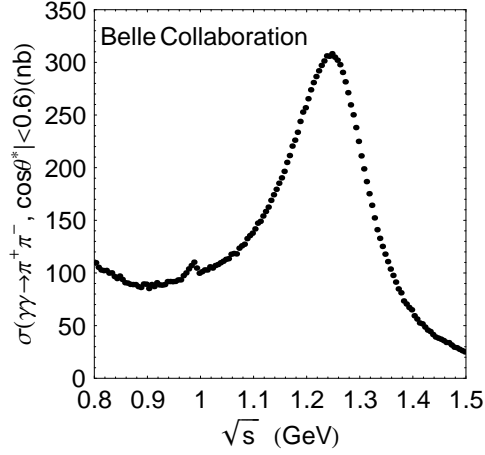}\vspace{-0.3cm}
\caption{{\footnotesize The high-statistics Belle data on $\gamma
\gamma\to\pi^+\pi^-$. A clear signal from the $f_0(980)$ has been
observed for the first time.}} \label{gg-pipi}
\end{center}\end{figure}

\begin{figure}\begin{center}
\includegraphics[height=60mm]{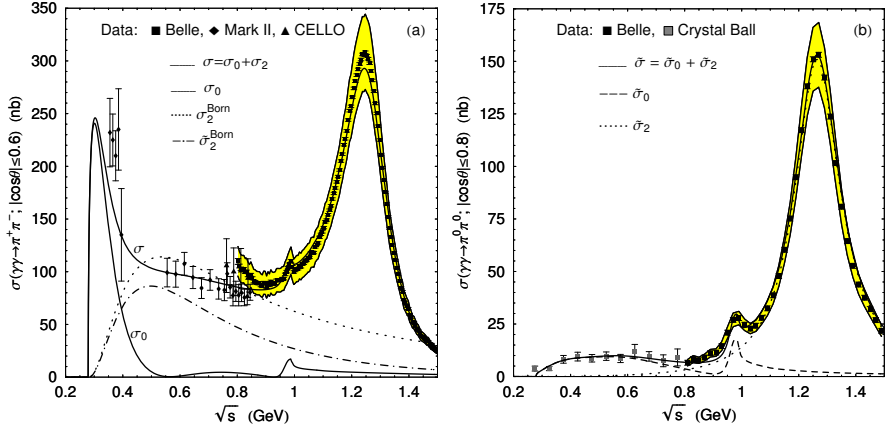}\vspace{-0.3cm} 
\caption{{\footnotesize The simultaneous description of the Belle
data on $\gamma\gamma\to\pi^+\pi^-$ and $\gamma\gamma\to\pi^0\pi^0
$. The bands show the size of the systematic errors of the Belle
data.}} \label{gg-pipi-pi0pi0}\end{center}\end{figure}

\begin{figure}\begin{center}
\includegraphics[height=60mm]{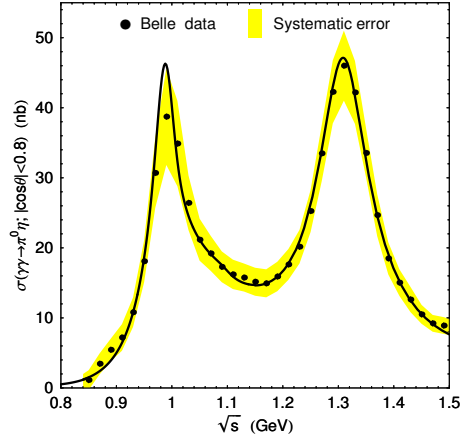} \vspace{-0.3cm}
\caption{{\footnotesize The description of the Belle data on
$\gamma\gamma\to\pi^0\eta$.}}\label{gg-pieta}\end{center}\end{figure}

Our statements for the $\sigma(600)$, $f_0(980)$, and $a_0(980)$
resonances are based on the detailed analysis of the new Belle
Collaboration data on the $\gamma\gamma$\,$\to$\,$\pi^+\pi^-$,
$\gamma\gamma$\,$\to$\,$ \pi^0\pi^0$, and $\gamma\gamma$\,$\to$\,$
\pi^0\eta$ reaction cross sections for energies up to 1.5 GeV. Owing
to huge statistics and high resolution in the invariant mass of the
$\pi\pi$ and $\pi^0\eta$ systems in the Belle experiments, clear
signals from the $f_0(980)$ and $a_0(980)$ resonances were detected.
The current experimental situation is shown in Figs. \ref{gg-pipi},
\ref{gg-pipi-pi0pi0}, and \ref{gg-pieta}.

{\boldmath \bf ii) \,Dynamics of the $\sigma(600)$ and $f_0(980)$
production in $\gamma\gamma\to\pi\pi$ \cite{AS07,AS05,AS08}.}

\begin{figure}\begin{center}\vspace{-0.4cm}
\includegraphics[width=65mm]{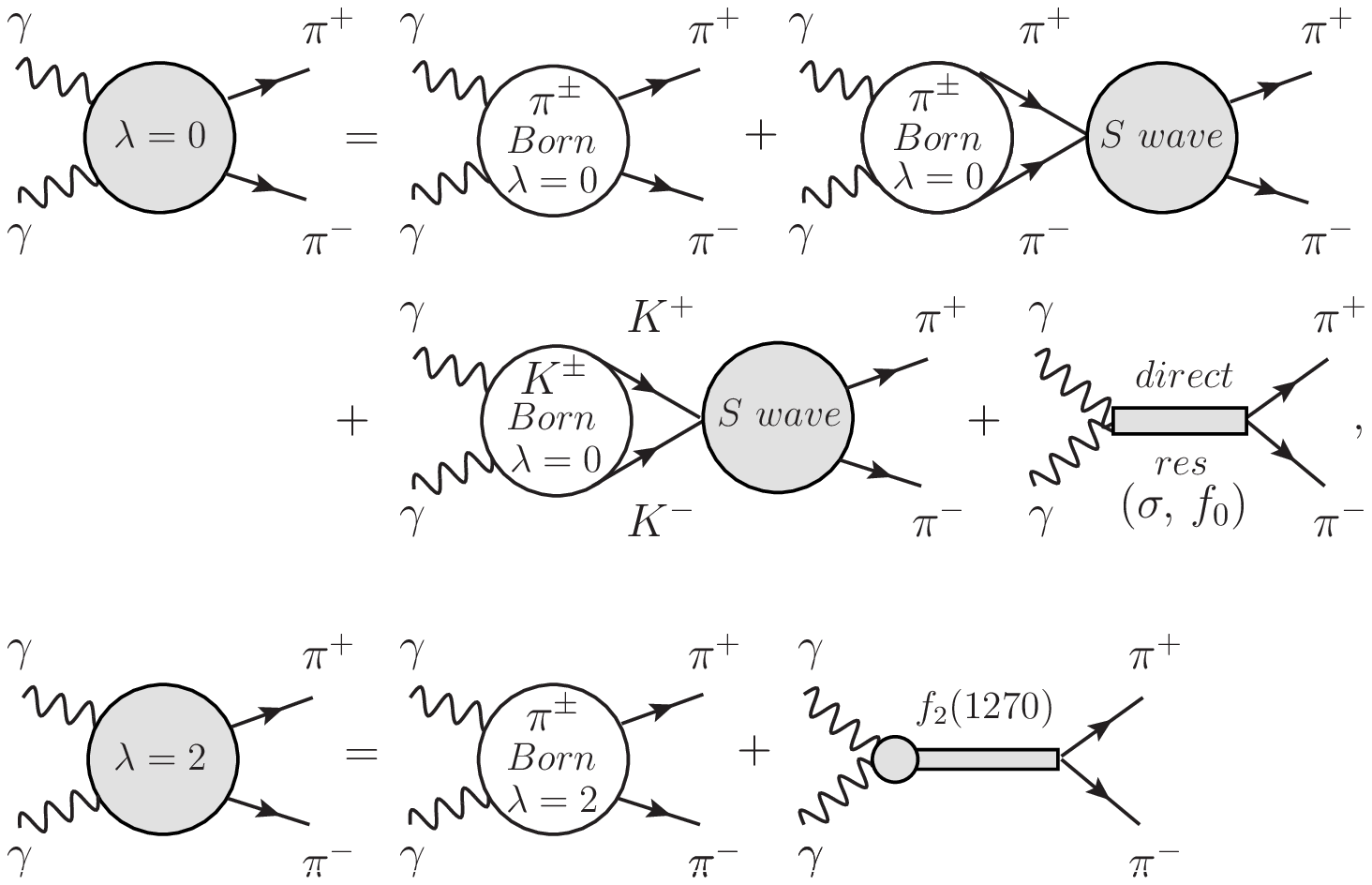} \hspace{0.4cm}
\includegraphics[width=65mm]{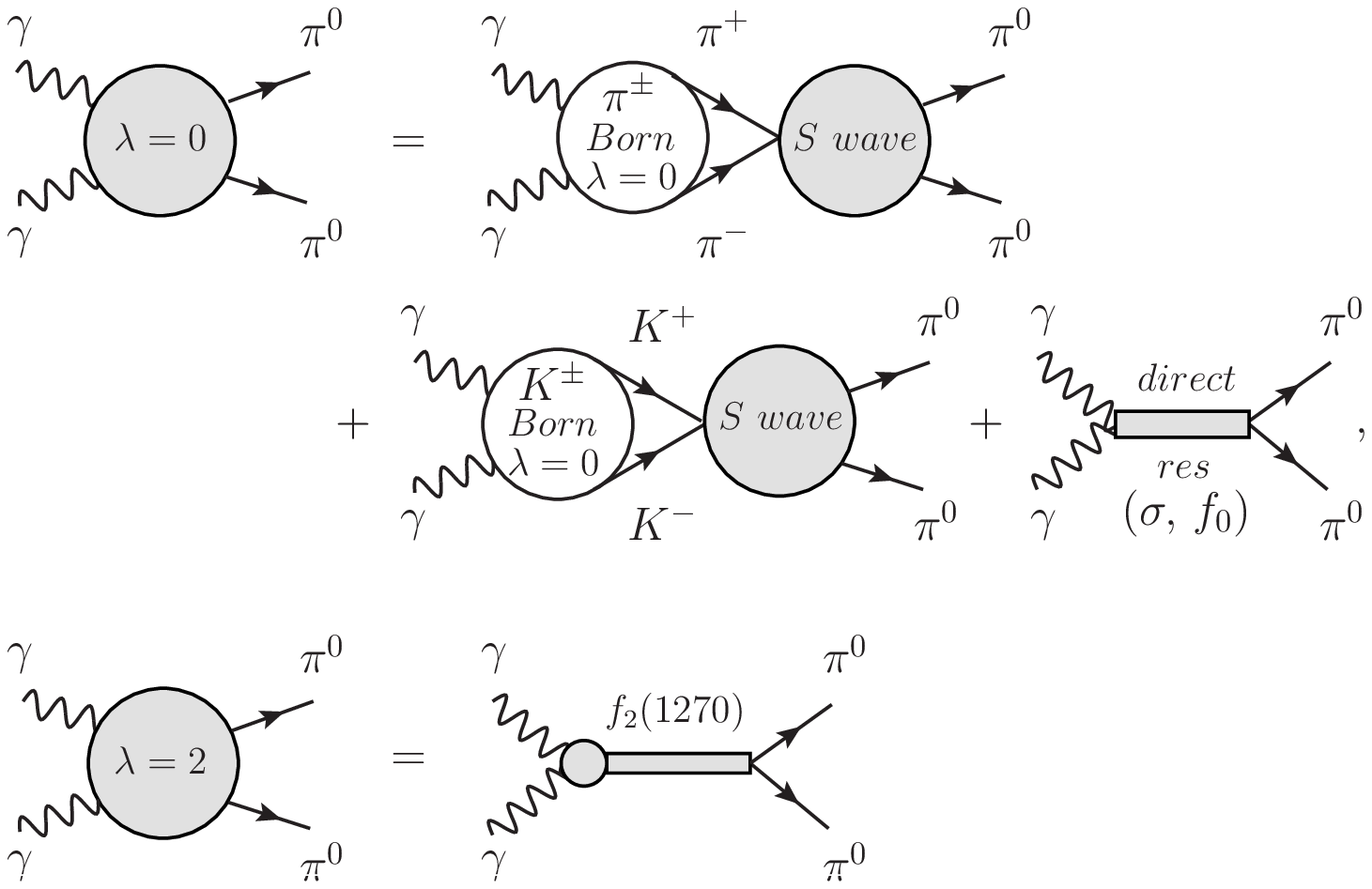}
\caption{{\footnotesize Dynamical model for the helicity amplitudes
$\gamma\gamma\to\pi^+\pi^-$ (left) and $\gamma\gamma\to\pi^0\pi^0$
(right).}} \label{Model-gg-pipi}\end{center}\end{figure}

\begin{figure}\begin{center}\vspace{-0.4cm}
\includegraphics[width=65mm]{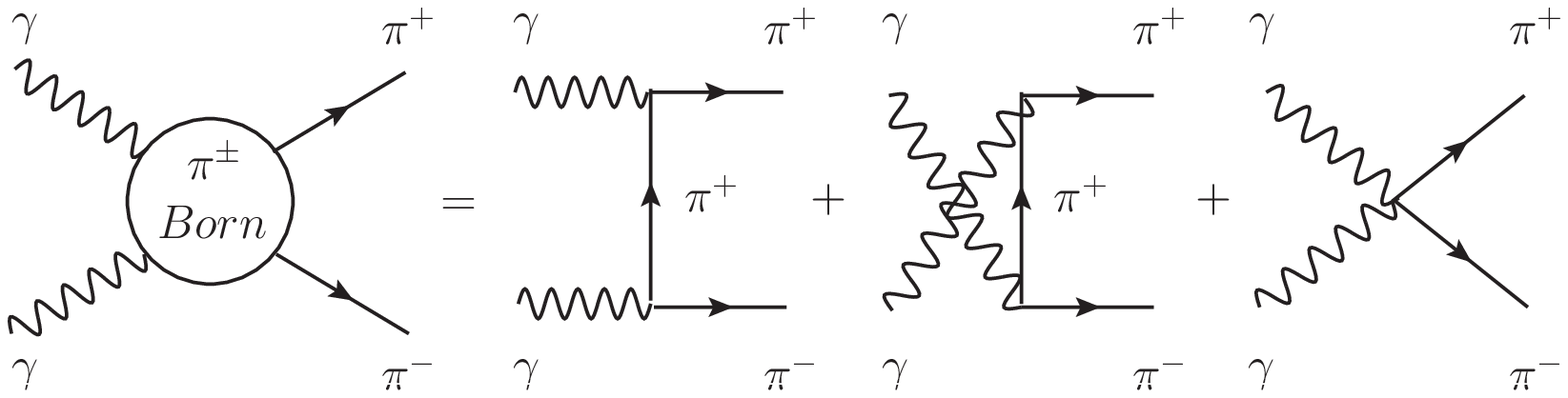} \hspace{0.4cm}
\includegraphics[width=65mm]{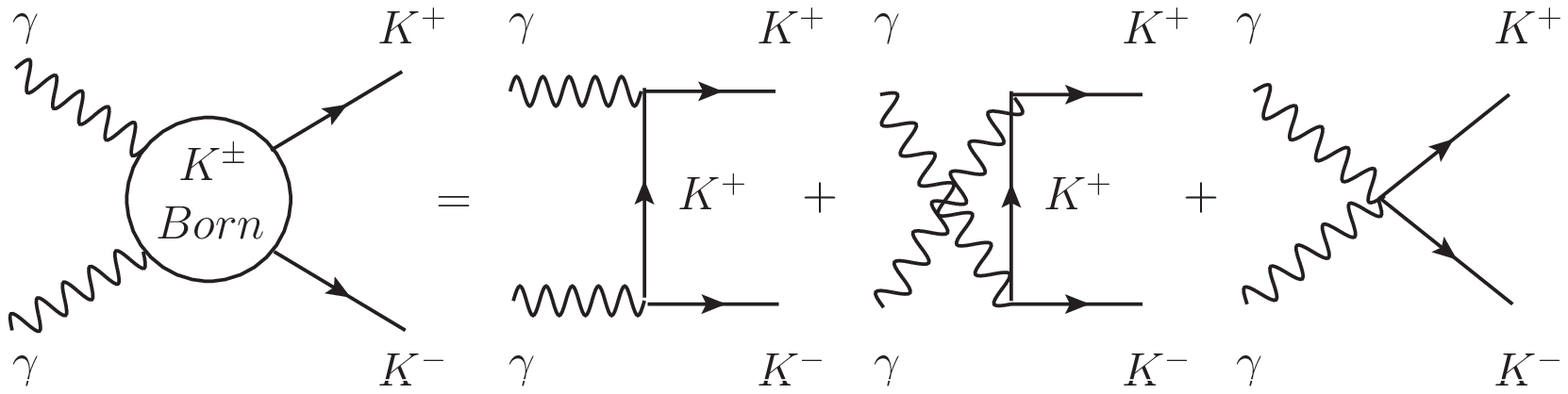}
\caption{{\footnotesize The Born one-pion (left) and one-kaon
(right) exchanges modified by form factors.}}
\label{Born-ope-oke}\end{center}\end{figure}

To analyze the data on the reactions $\gamma\gamma$\,$\to$\,$\pi^+
\pi^-$ and $\gamma\gamma$\,$\to$\,$\pi^0\pi^0$, we use a model for
the helicity and corresponding partial amplitudes, where the
electromagnetic Born terms caused by the one-pion and one-kaon
exchanges modified by form factors and strong elastic and inelastic
final-state interactions in $\pi^+\pi^-$ and $K^+K^-$ channels, as
well as the contributions due to the direct interaction of the
resonances with photons, are taken into account (Figs.
\ref{Model-gg-pipi}, \ref{Born-ope-oke}). Thus, symbolically,
Amplitude $=$ Born $+$ $\Sigma$ Born\,$\times$\,Strong\,FSI $+$
Direct. The amplitudes with definite isospin satisfy the Watson
theorem in the elastic region. The obtained simultaneous description
of the Belle data on $\gamma\gamma\to\pi^+\pi^-$ and
$\gamma\gamma\to\pi^0\pi^0$ is shown in Fig. \ref{gg-pipi-pi0pi0}
\cite{AS08}. The rescattering production mechanisms of the
$\sigma(600)$ and $f_0(980)$ resonances, i.e., the $q^2\bar q^2$
transitions, $\gamma\gamma\to\pi^+\pi^-\to\sigma(600)$ and
$\gamma\gamma\to K^+K^-\to f_0(980))$ dominate and indicate the
$q^2\bar q^2$ structure of these states.

We note also the recent analyses relevant to the Belle data on the
reactions $\gamma\gamma\to\pi\pi$ \cite{PMUW08,MWZZZ09,MNW10}.


{\boldmath \bf iii) \,Dynamics of the $a_0(980)$ production in
$\gamma\gamma\to\pi^0\eta$ \cite{AS88,AS10}.}

\begin{figure}\begin{center}
\includegraphics[width=65mm]{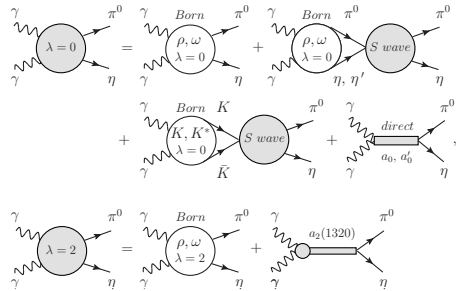}
\caption{{\footnotesize Dynamical model for the helicity amplitudes
$\gamma\gamma$\,$\to$\,$\pi^0\eta$.}}\label{Model-gg-pieta}
\end{center}\end{figure}

\begin{figure}\begin{center}
\includegraphics[width=55mm]{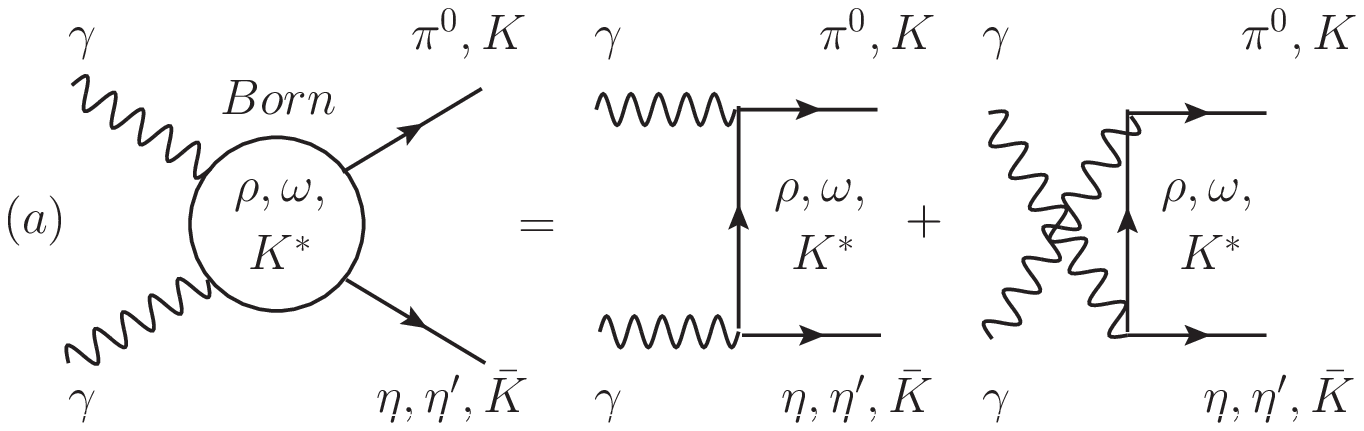}\hspace{0.4cm}
\includegraphics[width=65mm]{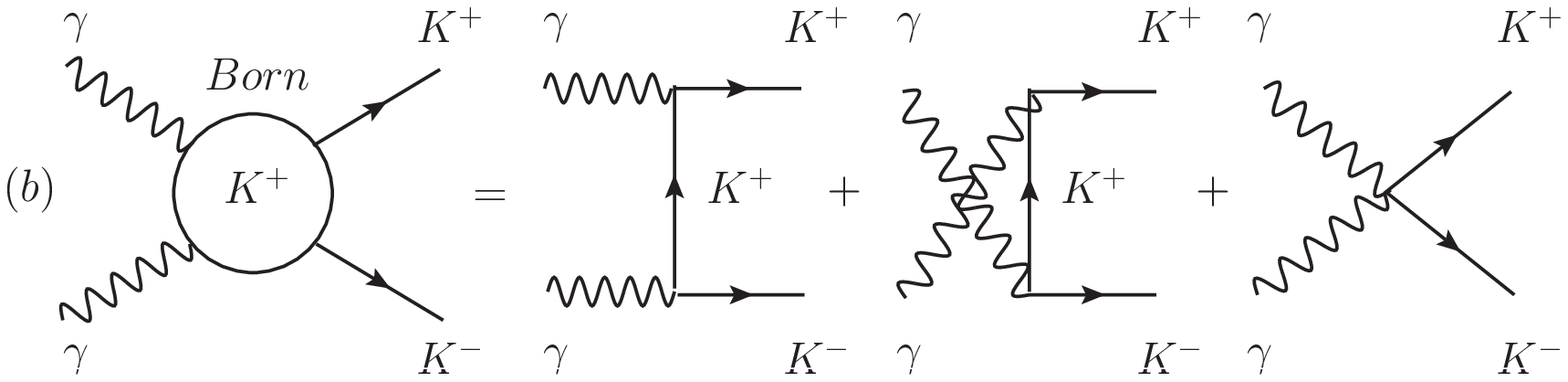}
\caption{{\footnotesize $(a)$ The $\rho$, $\omega$, $K^*$, and $(b)$
$K$ Born exchanges modified by form factors.}}\label{Born-gg-pieta}
\end{center}\end{figure}

Recently, we performed the analysis of the Belle data on the
reaction $\gamma\gamma\to\pi^0\eta$ \cite{AS10}. To do this, we have
significantly developed the model proposed previously in Ref.
\cite{AS88}. Figures \ref{Model-gg-pieta}, \ref{Born-gg-pieta} and
\ref{gg-pieta} illustrate our model and the resulting description of
the data on the $\gamma\gamma\to\pi^0\eta$ reaction cross section.
The experimentally observed pattern is the result of the combination
of many dynamical factors. The rescattering contributions are the
most essential ones.

The main constituents of the $\gamma\gamma$\,$\to$\,$\pi^0\eta$
reaction mechanism are the following. The inelastic rescattering
$\gamma\gamma$\,$\to$\,$K^+K^-$\,$\to$\,$\pi^0\eta$ with $K^+K^-$
produced via the one-kaon exchange mechanism specifies the natural
scale for the $a_0(980)$ production cross section in
$\gamma\gamma$\,$\to$\,$\pi^0\eta$. An estimate gives
$\sigma(\gamma\gamma$\,$\to$\,$K^+K^-$\,$\to$\,$ a_0(980)$\,$
\to$\,$\pi^0\eta;|\cos\theta|\leq0.8)\approx0.8 \cdot1.4\alpha^2
R_{a_0}/m^2_{a_0}\approx24$\,nb\,$\cdot R_{a_0}$ in the maximum,
where $R_{a_0}$\,=\,$ g^2_{a_0K^+K^-}/g^2_{a_0 \pi\eta}$. There is
the noticeable additional narrowing of the $a_0(980)$ peak due to
this mechanism in the $\gamma\gamma\to\pi^0\eta$ channel. The $K^*$
exchange narrows slightly the $a_0(980)$ peak too. However, the
$\gamma\gamma $\,$\to$\,$ K\bar K$\,$\to$\,$\pi^0\eta$ rescattering
mechanism alone cannot describe the data in the $a_0(980)$ resonance
region. The observed cross section can be obtain by adding the Born
$\rho$ and $\omega$ exchange contribution, modified by the $S$ wave
rescattering $\gamma \gamma$\,$\to$\,$(\pi^0\eta+\pi^0\eta')$\,$
\to$\,$\pi^0\eta$, and the amplitude caused by the direct
transitions of the $a_0(980)$ and heavy $a'_0$ resonances into
photons. Each of the contributions of these two mechanisms are not
too large in the $a_0(980)$ region. But the main thing is that their
coherent sum with the contribution of the inelastic rescattering
$\gamma\gamma$\,$\to$\,$ K\bar K$\,$\to$\,$\pi^0\eta$ leads to the
considerable enhancement of the $a_0(980)$ resonance manifestation.

\begin{figure}\begin{center}
\includegraphics[height=60mm]{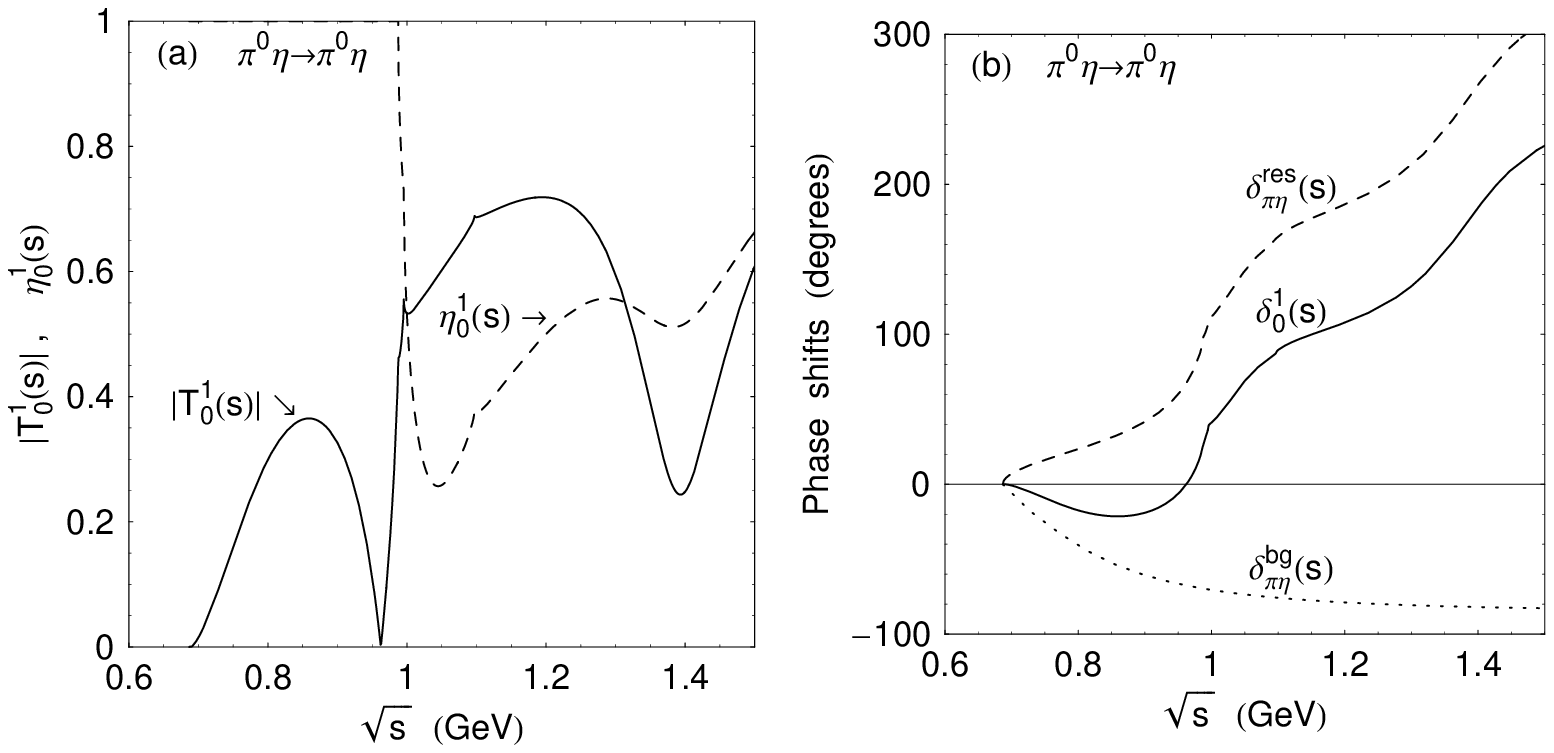} \vspace{-0.3cm}
\caption{{\footnotesize Modulus of $T^1_0(s)$, inelasticity
$\eta^1_0(s)$ (a), and phase shifts (b) of the $S$ wave amplitude
$\pi^0\eta\to\pi^0\eta$ (here the $\pi\eta$ scattering length
$a^1_0$\,=\,$0.0098m^{-1}_\pi$ is in agreement with the chiral
theory expectations $(0.005-0.01)m^{-1}_\pi$).}}
\label{pieta-pieta}\end{center}\end{figure}

One of the results of our analysis consists in the preliminary
information obtained on the $S$ wave amplitude of the reaction
$\pi^0\eta$\,$\to$\,$\pi^0\eta$ (Fig. \ref{pieta-pieta}), which is
important for the low-energy physics of pseudoscalar mesons.\\

{\boldmath \bf 4. Future trends: the $\sigma(600)$, $f_0(980)$, and
$a_0(980)$ investigations in $\gamma\gamma\to K\bar K$
$\mbox{\quad\,}$ and in $\gamma\gamma^*$ collisions \cite{AS10}.}

The Belle Collaboration has investigated the
$\gamma\gamma$\,$\to$\,$\pi^+\pi^-$, $\gamma\gamma$\,$\to$\,$
\pi^0\pi^0$, and $\gamma\gamma$\,$\to$\,$\pi^0\eta$ reactions with
the highest statistics. However, similar information is still
lacking for the processes $\gamma\gamma$\,$\to$\,$K^+K^-$ and
$\gamma\gamma$\,$\to $\,$K^0\bar K^0$. The $S$ wave contributions
from the $K^+K^-$ Born term, $f_0(980)$, and $a_0(980)$ resonances
near thresholds of these two channels are not clearly understood.
They can be measured with the Belle, L3, CLEO, and KLOE-2
detectors.

There are also the promising possibility of investigating the nature
of the light scalars in $\gamma\gamma^*$ collisions. If the
$\sigma(600)$, $f_0(980)$, and $a_0(980)$ are $q^2\bar q^2$ states,
their contributions to the $\gamma\gamma^*$\,$\to$\,$\pi\pi$ and
$\gamma\gamma^*$\,$\to $ $\pi^0\eta$ cross sections should decrease
with increasing $Q^2$ more rapidly than the contributions from the
classical tensor mesons $f_2(1270)$ and $a_2(1320)$. A similar
behavior of the contribution from the $q^2\bar q^2$ exotic resonance
state with $I^G=2^+$ and $J^{PC}=2^{++}$ to the
$\gamma\gamma^*$\,$\to$\,$\rho^0\rho^0$ and $\gamma
\gamma^*$\,$\to$\,$\rho^+ \rho^-$ cross sections was recently
observed by the L3 Collaboration \cite{L3}.\\

{\boldmath \bf 5. Summary.}\vspace{-0.1cm}

\begin{itemize}
\item
The mass spectrum of the light scalars, $\sigma (600)$, $\kappa
(800)$, $f_0(980)$, $a_0(980)$, gives an idea of their $q^2\bar q^2$
structure.\vspace{-0.1cm}
\item Both intensity and mechanism of the $a_0(980)/f_0(980)$ production
in the radiative decays of $\phi(1020)$, the  $q^2\bar q^2$
transitions $\phi\to K^+K^-\to\gamma [a_0(980)/f_0(980)]$, indicate
their $q^2\bar q^2$ nature.\vspace{-0.1cm}
\item Both intensity and mechanism of the scalar
meson decays into $\gamma\gamma$, namely, the $q^2\bar q^2$
transitions $\sigma(600)\to\pi^+\pi^-\to\gamma\gamma$,
$f_0(980)/a_0(980)$\,$\to$\,$K^+K^-$\,$\to$\,$\gamma\gamma$, and
$a_0(980)$\,$\to$\,$\pi^0\eta$\,$\to$\,$\gamma\gamma$ indicate their
$q^2\bar q^2$ nature also.\vspace{-0.1cm}
\item In addition, the absence of
$J/\psi$ $\to$ $\gamma f_0(980)$, $\rho a_0(980)$, $\omega f_0(980)$
in contrast to the intensive $J/\psi$ $\to$ $\gamma f_2(1270)$, $\
\gamma f'_2(1525)$, $\rho a_2(1320)$, $\omega f_2(1270)$ decays
intrigues against the $P$ wave $q\bar q$ structure of the $a_0(980)$
and $f_0(980)$.\vspace{-0.1cm}
\item It seems also undisputed that in all respects the $a_0(980)$ and
$f_0(980)$ mesons are strangers in the company of the well
established $b_1(1235)$, $h_1(1170)$, $a_1(1260)$, $f_1(1285)$,
$a_2(1320)$, and $f_2(1270)$ mesons which are the members of the
lower $P$ wave $q\bar q$ multiplet.
\end{itemize}

This work was supported in part by the RFFI Grant No. 10-02-00016
from the Russian Foundation for Basic Research.






\end{document}